\documentclass[notitlepage,twocolumn,draft]{article}
\usepackage{graphicx}
\usepackage{amsmath}
\usepackage{titling}
\usepackage{ulem}
\usepackage[colorlinks]{hyperref}
\usepackage[numbers]{natbib}
\newcommand{\dd}{\ensuremath{\text{d}}}
\newcommand{\first}{\ensuremath{1^\text{st}}}
\newcommand{\second}{\ensuremath{2^\text{nd}}}
\pretitle{\begin{center}\Huge\bfseries}
\posttitle{\par\end{center}\vskip 0.5em}
\preauthor{\begin{center}\Large\ttfamily}
\postauthor{\end{center}}
\predate{\par\large\centering}
\postdate{\par}

\begin{document}
\title{Quantum-mechanical treatment of two particles in a potential box}
\author{Gottfried Mann\thanks{Leibniz-Institut f\"{u}r Astrophysik Potsdam (AIP), An der Sternwarte 16, D-14482 Potsdam, Germany, e-mail: gmann@aip.de}}
\date{}
\maketitle
\begin{abstract}
In classical physics, there is a basic principle, namely ``A particle cannot be located at the position of another one on the same time''.
Which consequences can be derived if this principle is transferred into quantum physics? For doing that, two distinguishable particles are considered to be trapped in a potential box by means of the Schr\"odinger equation. In result,
the particles can necessarily be located only at discrete positions.
\end{abstract}

\section{Introduction}
\label{sec:1}
In classical physics, there is a basic principle, which is called ``Where a body is located, there cannot be another one at the same time.'' (see e.g. \cite{meschede15} as a textbook). In the present paper, it is discussed what results from this principle, if it is transferred into quantum physics.
For doing that, two distinguishable, impenetrable, pointlike particles trapped in a potential box are considered in terms of quantum mechanics. For this case, the stationary one-dimensional Schr\"odinger equation (see e.g. \cite{landau74} as a textbook) can be written as
\begin{eqnarray}
\label{eqn:01}
E \cdot \psi(x_{1},x_{2}) & = & - \frac{\hbar^{2}}{m_{1}^{2}} \cdot \frac{\partial^{2}\psi}{\partial x_{1}^{2}} 
                   - \frac{\hbar^{2}}{m_{2}^{2}} \cdot \frac{\partial^{2}\psi}{\partial x_{2}^{2}} \nonumber \\
             &   & + V_\text{box}(x_{1}) \cdot \psi + V_\text{box}(x_{2})\cdot \psi \nonumber \\
             &   & + V_\text{int}(x_{1},x_{2}) \cdot \psi
\end{eqnarray}
($\hbar = h/2\pi$ with Planck's constant $h$) with $E$ as the energy of the two-particle system. $m_{1}$ and $m_{2}$ denote the masses of the \first{} and \second{} particle,
respectively. The wave function $\psi(x_{1},x_{2})$ is depending on the spatial coordinates $x_{1}$ and $x_{2}$
of the \first{} and \second{} particle, respectively. $V_\text{box}(x)$ describes the potential of the potential box with the spatial length $L$:
\begin{eqnarray}
\label{eqn:02}
 V_\text{box}(x) &=& \begin{cases}
                      0 & \text{for} \quad 0 < x < L \\
                 \infty & \text{otherwise}
             \end{cases}
\end{eqnarray}
(Here, $x$ should be taken either for $x_{1}$ or for $x_{2}$.) As argued above, one particle cannot be at the position of the other one on the same time.
That can be regarded as some kind of interaction, which is described by the potential
\footnote{Note that the definition of the potential $V_\text{int}$ should not be confused
with the well-known Dirac's delta-function as discussed in footnote 2.}
\begin{eqnarray}
\label{eqn:03}
V_\text{int}(x_{1},x_{2})&=& \begin{cases}
                      \infty & \text{for} \quad x_{1} = x_{2}\\
                           0 & \text{for} \quad x_{1} \not= x_{2}
             \end{cases}
\end{eqnarray}
In quantum mechanics, $\|\psi(x_{1},x_{2})\|^{2} \cdot \dd{}x_{1}\dd{}x_{2}$ 
gives the probability, that the \first{} and \second{} particle are located in the intervals $(x_{1},x_{1}+\dd{}x_{1})$ and $(x_{2},x_{2}+\dd{}x_{2})$
(see e.~g. \cite{landau74}). Because of this meaning of the wave function, the inclusion of the potential $V_{int}$ into Eq.~(\ref{eqn:01})
leads to the requirement, that the wave function must vanish for $x_{1} = x_{2}$. (see for more details Appendix~\ref{sec:6}). There, a particle is considered in a potential box in which an infinitely thin and high potential wall is additionally inserted at the position x = 0. It is shown, that the wave function has consequently a zero there.

Thus, the task to be discussed is completely defined by 
Eqs.~(\ref{eqn:01}, \ref{eqn:02}, \ref{eqn:03}). 
\section{Classical solution in quantum mechanics}
\label{sec:2}
The treatment of a particle in the potential box is a classical task in textbooks of quantum mechanics (see \cite{landau74,dawydow78} and also Appendix~\ref{sec:6}). Now, the case in which both particles are trapped in the potential box without any interaction, is considered, i.e. the potential $V_\text{int}$ is neglected. 
Then, Eq.~(\ref{eqn:01}) reduces to
\begin{eqnarray}
\label{eqn:04}
E \cdot \psi(x_{1},x_{2}) & = & - \frac{\hbar^{2}}{m_{1}^{2}} \cdot \frac{\partial^{2}\psi}{\partial x_{1}^{2}} 
                   - \frac{\hbar^{2}}{m_{2}^{2}} \cdot \frac{\partial^{2}\psi}{\partial x_{2}^{2}} \nonumber \\
             &   & + V_\text{box}(x_{1}) \cdot \psi + V_\text{box}(x_{2})\cdot \psi \nonumber \\
\end{eqnarray}
The potential $V_\text{box}$ requires that the wave function $\psi(x_{1}, x_{2})$ vanishes at the walls of the potential box and outside of the  region $0 < x_{1}, x_{2} < L$. Then, the solution of Eq.~(\ref{eqn:04}) is found to be
\begin{equation}
\label{eqn:05}
\psi(x_{1}, x_{2}) = \frac{2}{L} \cdot \sin(k_{1}x_{1}) \cdot \sin(k_{2}x_{2})
\end{equation}
with
\begin{equation}
\label{eqn:06}
k_{1} = \frac{\sqrt{2m_{1}E_{1}}}{\hbar} = n_{1} \cdot \frac{\pi}{L}
\end{equation}
\begin{equation}
\label{eqn:07}
k_{2} = \frac{\sqrt{2m_{2}E_{2}}}{\hbar} = n_{2} \cdot \frac{\pi}{L}
\end{equation}
($n_{1}, n_{2} = 1, 2, 3, \dots$) and 
\begin{equation}
\label{eqn:08}
E = E_{1} + E_{2} 
  = \frac{\hbar^{2}k_{1}^{2}}{2m_{1}} + \frac{\hbar^{2}k_{2}^{2}}{2m_{2}}
\end{equation}
in the region  $0 < x_{1}, x_{2} < L$. Note, that the wave function is normalized to unity and is a real one, here. $E_{1}$ and $E_{2}$ denote the energies of the \first{} and \second{} particle, respectively. 
\section{Inclusion of the potential $V_\text{int}$}
\label{sec:3}
The wave function (Eq.~(\ref{eqn:05})) is adopted for describing
both particles in the potential box. 
The potential $V_\text{int}(x_{1},x_{2})$ (see Eq.~(\ref{eqn:03})) describes the interaction of both particles with each other. As discussed in Section~\ref{sec:1}, the wave function (Eq.~(\ref{eqn:05})) must have 
a zero at $x_{1} = x_{2}$ because of the potential $V_{int}$ leading to
\begin{equation}
\label{eqn:09}
\cos[(k_{1}-k_{2})x_{2}] = \cos[(k_{1}+k_{2})x_{2}]
\end{equation}
This equation is fulfilled for $\beta = 2\pi\nu_{1} + \alpha$ and/or $\beta = 2\pi\nu_{2} - \alpha$
with $\alpha = (k_{1}-k_{2})x_{2}$ and $\beta = (k_{1}+k_{2})x_{2}$ and 
$\nu_{1},\nu_{2} = 1, 2, 3, \dots$, leading to
$\alpha + \beta = 2k_{1}x_{2} = 2\pi\nu_{2}$ and
\begin{equation}
\label{eqn:10}
x_{2} = \frac{\nu_{2}\pi}{k_{1}} = \frac{\nu_{2}}{n_{1}} \cdot L
\end{equation}
with $\nu_{2} \leq n_{1}$ because of $x_{2} \leq L$.
(Here, Eq.~(\ref{eqn:06}) has been used.)
For $x_{2} \rightarrow x_{1}$, the same procedure provides
\begin{equation}
\label{eqn:11}
x_{1} = \frac{\nu_{1}\pi}{k_{2}} = \frac{\nu_{1}}{n_{2}} \cdot L
\end{equation}
with $\nu_{1} \leq n_{2}$. 

According to Eqs.~(\ref{eqn:10}) and (\ref{eqn:11}), the nearest positions 
of the \first{} and \second{} particle to the potential wall at $x = L$
are $x_{2} = [(n_{1}-1)/n_{1}]L$ and $x_{1} = [(n_{2}-1)/n_{2}]L$.
Then, the distance between these positions are found to be
\begin{equation}
\label{eqn:12}
\frac{\|x_{1}-x_{2}\|}{L}   
                      = \frac{\|n_{2}-n_{1}\|}{n_{1}n_{2}} \cdot \frac{1}{L}
\end{equation} 
Hence, the distance of two neighbouring positions of the \first{} and 
\second{} particle becomes smaller for  $n_{1}, n_{2} \rightarrow \infty$ but never exactly zero as it should be, ´taking into account $n_{1} \not= n_{2}$. 
Note, that, as well-known, quantum-mechanical results tranfers into the classical ones for large quantum numbers, i. e. $n_{1}, n_{2} \rightarrow \infty$. In the regime of classical mechanics, all positions in the 
box have the same probability to be occupied by the particles  as discussed
in Appendix~\ref{sec:7}.

Inserting Eqs.~(\ref{eqn:10}) and (\ref{eqn:11}) into Eq.~(\ref{eqn:05}), the wave function has the from
\begin{equation}
\label{eqn:13}
\psi(x_{1},x_{2}) = \frac{4}{L^{2}} \cdot \sin \left ( \frac{\nu_{1,i}}{n_{2}} n_{1} \pi \right) \cdot
                     \sin \left ( \frac{\nu_{2,j}}{n_{1}} n_{2} \pi \right)
\end{equation}
with $\nu_{1,i} = 1, \dots , n_{2}$ and $\nu_{2,j} = 1, \dots , n_{1}$. Now, the wave function $\psi$ is actually not a function but a matrix. 
According to the interpretation of the wave function in quantum mechanics,
the matrix $W_{i,j} \propto \tilde W_{i,j} = \|\psi\|^{2}$ gives the probability, that the \first{} and \second{} particle are located at the $i$-th and $j$-th 
position. That leads to 
\begin{eqnarray}
\label{eqn:14}
\tilde W_{i,j} &=& \frac{4}{L^{2}} \cdot \sin^{2} \left( \frac{\nu_{1}}{n_{2}} n_{1} \pi \right) \nonumber\\
                               & & \cdot \sin^{2} \left( \frac{\nu_{2}}{n_{1}} n_{2} \pi \right)
\end{eqnarray}
Since $W_{i,j}$ should be normalized to unity, i.e.
\begin{equation}
\label{eqn:15}
\sum_{i,j} \tilde W_{i,j} = 1
\end{equation}
one finds for 
\begin{equation}
\label{eqn:16}
W_{i,j} = \frac{\tilde W_{i,j}}{\sum_{i,j} \tilde W_{i,j}}
\end{equation}
Now, the matrix $W_{i,j}$ is completely determined
by Eqs.~(\ref{eqn:14}), (\ref{eqn:15}), and (\ref{eqn:16}).

In result, the appearance of a second particle in the potential box and the principle, that a particle cannot be at a place of the other one at the same time, leads consequently to the fact, that both particles can only be located at discrete positions in the potential box with individual probabilities, which are always smaller than unity.
\section{An example}
\label{sec:4}
For illustrating the procedure presented in the previous Section, 
a special example will be discussed. Low quantum states are chosen,
namely $n_{1} = 3$ and $n_{2} = 4$, for instance. Hence, $3$ and $2$ positions are disposable for the \first{} and \second{} particle, respectively, i.e.
\begin{equation}
\label{eqn:17}
x_{11} = \frac{L}{4} \qquad  x_{12} = \frac{L}{2} \qquad x_{13} = \frac{3L}{4}
\end{equation}
and
\begin{equation}
\label{eqn:18}
x_{21} = \frac{L}{3} \qquad  x_{22} = \frac{2L}{3} 
\end{equation}
$x_{1i}$ and $x_{2j}$ denote the positions of the \first{} and \second{} particle, respectively.
$\tilde W_{ij} = \|\psi(x_{1i},x_{2j})\|^{2}$ are the relative probabilities, that the \first{} and \second{} particle take the $i$-th and $j$-th positions. Hence, one gets $2 \times 3$ elements for the matrix
%
\begin{equation}
\label{eqn:19}
\tilde W_{ij} = \frac{4}{L^{2}} \cdot \sin^{2} \left (\frac{\nu_{1i}}{4} \cdot 3\pi \right ) \cdot
                \sin^{2} \left (\frac{\nu_{2j}}{3} \cdot 4\pi \right ) 
\end{equation}
with $\nu_{1i} = 1, 2, 3$ and $\nu_{2j} = 1, 2$. For instance, one finds for 
$\tilde W_{1,2} \propto \sin^{2}(3\pi/4) \cdot \sin^{2}(8\pi/3) = 3/8$.
Following this procedure and taking into account Eqs.~(\ref{eqn:15}),
(\ref{eqn:17}), and (\ref{eqn:18}), the matrix of probability (see Eq.~(\ref{eqn:16})) is found to be
\begin{eqnarray}
\label{eqn:20}
{W_{ij}} &=& 
\left ( \begin{array}{cc} 
1/8 & 1/8  \\
1/4 & 1/4 \\
1/8 & 1/8  
\end{array} \right )
\end{eqnarray}

With the knowledge of the matrix $W_{ij}$, the mean positions of the \first{} and \second{} particle within the potential box can be calculated to be
$\bar x_{1} = L/2$ and $\bar x_{2} = L/2$, respectively, according to
\begin{eqnarray}
\label{eqn:21}
\!\! \bar x_{1} = \sum_{i,j} x_{1i} \cdot W_{ij} &\,\text{and}\,& \bar x_{2} = \sum_{i,j} x_{2i} \cdot W_{ij}
\end{eqnarray}
Both values agree with the value of the classical approach (see 
Eq.~(\ref{eqn:49}))
as presumed. The mean square $\overline{(\Delta x_{1})^{2}}$ of the fluctuations 
of the positions of the \first{} particle around its mean value $\bar x_{1}$
is defined by
\begin{equation}
\label{eqn:22}
\overline{(\Delta x_{1})^{2}} = \overline{x_{1}^{2}} - (\bar x_{1})^{2}
\end{equation} 
with
\begin{equation}
\label{eqn:23}
\overline{x_{1}^{2}} = \sum_{i,j} x_{1i}^{2} \cdot W_{ij}
\end{equation}
Following this procedure, one finds 
$\overline{(\Delta x_{1})^{2}}/L^{2}$ = 1/32 and
$\overline{(\Delta x_{2})^{2}}/L^{2}$ = 1/36 
for the \first{} and \second{} particle, respectively.
Both values are smaller than the value of 1/12 
(see Eq.~(\ref{eqn:52})) in the classical approach. 

The distances of neighbouring positions of the \first{} and \second{} particle
are $L/12$, $5L/12$, and/or $L/6$ taking into account Eqs.~(\ref{eqn:17}), and (\ref{eqn:18}).
The mean distance $\bar d$ between both particles is defined by
\begin{equation}
\label{eqn:24}
\bar d = \sum_{i,j}\ \|x_{1i} - x_{2j}\| \cdot W_{ij}
\end{equation}
and is calculated to be $\bar d/L = 5/24 (\approx 0.2083)$. Note, that the classical approach
provides for $\bar d/L = 1/3$ (see Eq.~(\ref{eqn:54})). 
Then, the mean square $\overline{(\Delta d)^{2}}/L^{2}$
of the fluctuation of the distance $d$ around the mean value $\bar d$
is found to be $\overline{(\Delta d)^{2}}/L^{2} = 1/64 (\approx 0.0156)$ by means of
\begin{equation}
\label{eqn:25}
\overline{(\Delta d)^{2}} = \overline{d^{2}} - (\bar d)^{2}
\end{equation}
and
\begin{equation}
\label{eqn:26}
\overline{d^{2}} = \sum_{i,j} \|x_{1i} - x_{2j}\|^{2} \cdot W_{ij}.
\end{equation}
This value is smaller than $1/18$, which is the value of the classical approach
(see Eq.~(\ref{eqn:56})).

In the case of discussion, the derived values of $\overline{(\Delta x_{1})^{2}}/L^{2}$, $\overline{(\Delta x_{2})^{2}}/L^{2}$, $\bar d$, and 
$\overline{(\Delta d)^{2}}/L^{2}$ are smaller than those of the classical approach
(see Appendix~\ref{sec:7}). That can be explained in the following manner:
As already mentioned, $\|\psi(x_{1},x_{2})\|^{2}$ gives the density of probability, that the \first{} and \second{}particle are located in the intervals $(x_{1}, x_{1}+\text{d}x_{1})$ and $(x_{2}, x_{2}+\text{d}x_{2})$
(see \cite{landau74} as a textbook). Since 
$\|\psi\|^{2} \propto \sin^{2}(k_{1}x_{1}) \cdot \sin^{2}(k_{2}x_{2})$
(see Eq.~(\ref{eqn:05})), this probability is greater in the middle of the box than at its edges. Thus, one can say, that the walls of the potential box acts like a repulsive force on the particle. That is a pure quantum-mechanical effect, since it does not appear in classical physics (see Appendix~\ref{sec:7}). Since both particles are 
more located at the middle of the box, consequently, their distance to each other
becomes also smaller in comparison to the classical approach (see Appendix~\ref{sec:7}).

In Section~\ref{sec:3}, it has been shown, that the particles are localized at discrete positions, This result seems to contradict to Heisenberg's inequality
(see \cite{landau74}). But it is not the case as demonstrated in the framework of the example discussed in this Section: In quantum mechanics, the momentum $p$ is given by $p = \hbar k$ (\cite{landau74}) with $k$ as the wave number.
In the case of a particle with the mass $m$ trapped in a potential box,
one finds $\bar p = 0$ and $\overline{(\Delta p)^{2}} = (\hbar k)^{2}$
(see \cite{dawydow78}). The \first{} particle has the quantum number $n_{1} = 3$ leading to $\overline{(\Delta p_{1})^{2}} = (9\hbar \pi/L)^{2}$ (see Eq.~(\ref{eqn:06})),
$\overline{(\Delta x_{1})^{2}} = 1/32$, and, finally, to
$\overline{(\Delta p)^{2}} \cdot \overline{(\Delta x_{1})^{2}} =
(81\pi^{2}\hbar^{2}/32) \geq \hbar^{2}/4$. 
A similar result is obtained for the \second{} particle by the same procedure.
Hence, the results presented in this paper does not contradict to Heisenberg's
inequality, as it should be.
\section{Summary}
\label{sec:5}
In this paper, the basic principle of classical physics, 
namely ``A particle cannot be located at the position of another one on the same time'' (see e.g. \cite{meschede15} as a textbook), is transferred into quantum mechanics. For doing that, two distinguishable particles with different masses trapped in a potential box are considered in the framework of quantum mechanics. This task is treated by means of the stationary one-dimensional Schr\"odinger equation. 

In Appendix~\ref{sec:6}, the problem of one particle in the potential box is studied with an additional insertion of an infinitely thin wall with an infinitely high potential at the position $x = 0$. In quantum mechanics, the particle can penetrate through the wall due to the ``tunnel effect'' \cite{dawydow78}, but its wave function must have necessarily a zero at $x = 0$ as shown in Appendix~\ref{sec:6}.

In order to describe that one particle cannot be located at the place of the other one at the same time, the potential $V_{int}$ (see Eq.~(\ref{eqn:03})) is introduced in Eq.~(\ref{eqn:01}). It is infinite at $x_{1} = x_{2}$ and zero otherwise as defined by Eq.~(\ref{eqn:03}). According to the result in Appendix~\ref{sec:6}, the wave function $\psi(x_{1},x_{2})$ must have a zero at $x_{1} = x_{2}$. Consequently, both particles must be located at individual discrete positions as discussed in Section~\ref{sec:3}. Hence, the wave function $\psi$ has to be substituted by a matrix $W_{i,j}$. This matrix $W_{i,j}$ gives the probability, that the \first{} and \second{} particle are located at the positions $x_{1,i}$ and $x_{2,j}$, respectively, as demonstrated in terms of an example in Section~\ref{sec:4}.

In order to avoid misunderstandings, it should be emphasized that the approach presented in this paper has nothing to do with ``fermions'' and ``bosons'', which are basically indistinguishable particles. Here, two particles with different masses, i.e. distinguishable particles, are considered to be trapped in a potential box. The addition of a second particle in the potential box leads inevitably to a discretization of the locations of the particles in the box if one takes into account that an individual particle cannot be be at the position of the other ones at the same time. It should be emphasized that the location of the particles at discrete positions does not violate Heisenberg's inequality as demonstrated in Section~\ref{sec:4}. 
They are localized at discrete positions with different probabilities, which are nowhere unity. 
\section*{Acknowledgments}
The author is grateful to Hakan \"O{}nel for his \LaTeX{} support during the preparation of this manuscript.
\appendix
\section{Quantenmechanical treatment of a single particle in a one-dimensional potential box with a potential wall} 
\label{sec:6}
A single particle with the mass $m$ and the energy $E$ is considered to be trapped in the potential box with the length $L$. Additionally, a potential wall of the width $b$ is inserted at $x = 0$.
%
%
Then, the stationary one-dimensional Schr\"odinger equation of this problem is written as
\begin{equation}
\label{eqn:27}
- \frac{\hbar^{2}}{2m} \cdot \frac{d^{2}\psi}{\dd{}x^{2}} + V_{box}\psi + V_{wall}\psi = E \cdot \psi(x)
\end{equation}
$V_{box}$ describes the potential of the box as given by 
\begin{eqnarray}
\label{eqn:28}
\!\!\! V_{box}(x) & = & \begin{cases}
                  0 & \text{for} \quad -L/2 < x < L \\
                  \infty & \text{otherwise}
                \end{cases}
\end{eqnarray}
The potential wall with the spatial width $b$ is inserted at $x = 0$. Its potential $V_{wall}$ is described by
\begin{eqnarray}
\label{eqn:29}
\!\!\! V_{wall}(x) & = & \begin{cases}
                  V_{0} & \text{for} \quad -b/2 < x < b/2 \\
                  0   & \text{otherwise}
                \end{cases}
\end{eqnarray}
Now, the inclusion of the potential wall at the position $x = 0$ within the box is considered. (Here, we follow the procedure presented in \cite{dawydow78} p. 94.)
Then, the wave function $\psi(x)$ has the general form

%
\begin{eqnarray}
\label{eqn:30}
\psi =  \begin{cases}
                  A\sin(kx) + B\cos(kx) \hspace{0.3cm} \text{at} 
                  -L/2 < x < -b/2 \\
                  Ce^{-\kappa x} + De^{\kappa x} \hspace{0.3cm} \text{at} 
                  -b/2 \leq x \leq b/2 \\
                  E\sin(kx) + F\cos(kx) \hspace{0.3cm} \text{at} 
                  +b/2 < x < L 
\end{cases}
\end{eqnarray}
%
with
\begin{equation}
\label{eqn:31}
k = \sqrt{2mE}/\hbar
\end{equation}
and
\begin{equation}
\label{eqn:32}
\kappa = \sqrt{2m(V_{0}-E})/\hbar
\end{equation}
Here, $V_{0} > E$ is generally assumed, so that $k$ and $\kappa$ are real and positive quantities.

The task to be treated is symmetrical with respect to $x = 0$. Consequently, 
the resulted wave function $\psi(x)$ is either symmetrical or antisymmetrical.
The symmetrical and antisymmetrical case are given by $A = E = 0$ and
$C = D$, and $B = F = 0$ and $C = - D$, respectively. Because of the shape of
$V_{box}$, the wave function $\psi$ must vanish at $x = \pm L/2$. That leads to
\begin{equation}
\label{eqn:33}
k_{n} = \frac{(2n-1)\pi}{L}  \quad \text{for} \quad n = 1; 2; 3; ...
\end{equation} 
and
\begin{equation}
\label{eqn:34}
k_{n} = \frac{2n\pi}{L}  \quad \text{for} \quad n = 1; 2; 3; ...
\end{equation} 
for the symmetrical and antisymmetrical case, respectively.

The wave function and its first derivation with respect to $x$ must be
continuous at $x = \pm b/2$ (see \cite{dawydow78}). In the symmetrical case, it leads to
\begin{equation}
\label{eqn:35}
B \cdot \cos(kb/2) = C \cdot \left (e^{\kappa b/2} + e^{-\kappa b/2} \right )
\end{equation}
and
\begin{equation}
\label{eqn:36}
kB \cdot \sin(kd/2) = -\kappa C \cdot \left (e^{\kappa d/2} - e^{-\kappa d/2} \right )
\end{equation}
at $x = -b/2$. Eqn.~(\ref{eqn:35}) and (\ref{eqn:36}) represent a homogeneous system
of equations with the determinant
\begin{equation}
\label{eqn:37}
\frac{k}{\kappa} \cdot \sin(kb/2) = -\cos(kb/2) \cdot 
   \frac{\left (e^{\kappa b/2} - e^{-\kappa b/2} \right )}
   {\left (e^{\kappa b/2} + e^{-\kappa b/2} \right )}
\end{equation}
As considered here, the potential wall has an infinitely high potential and
an infinitely thin width. It means $k/\kappa \ll 1$ 
(because of $V_{0} \gg E$) and $kb \ll 1$. Under these conditions, the left 
hand side of Eq.~(\ref{eqn:37}) vanishes whereas the right hand site of it
does it only for $\kappa b = 0$.
Hence, there are no symmetrical solutions of the 
task for arbitrary values of $\kappa b$. 
In the antisymmetrical case, Eqn.~(\ref{eqn:38}) and (\ref{eqn:39})
\begin{equation}
\label{eqn:38}
-A \cdot \sin(kb/2) = C \cdot \left (e^{\kappa b/2} - e^{-\kappa b/2} \right )
\end{equation}
and
\begin{equation}
\label{eqn:39}
kA \cdot \cos(kb/2) = -\kappa C \cdot \left (e^{\kappa b/2} + e^{-\kappa b/2} \right )
\end{equation}
results from the requirement of continuity of the wave function and its
first derivative with respect to $x$ at $x = -b/2$
\begin{equation}
\label{eqn:40}
B \cdot \cos(kb/2) = C \cdot \left (e^{\kappa b/2} + e^{-\kappa b/2} \right )
\end{equation}
and
\begin{equation}
\label{eqn:41}
kB \cdot \sin(kd/2) = -\kappa C \cdot \left (e^{\kappa b/2} - e^{-\kappa b/2} \right )
\end{equation}
at $x = -b/2$.
The determinant of this homogeneous system of equations (see Eqn.~(\ref{eqn:40}) and (\ref{eqn:41})) is found to be 
\begin{equation}
\label{eqn:42}
\frac{k}{\kappa} \cdot \cos(kd/2) = \sin(kd/2) \cdot 
   \frac{\left (e^{\kappa d/2} + e^{-\kappa d/2} \right )}
   {\left (e^{\kappa d/2} - e^{-\kappa d/2} \right )}
\end{equation}
The denominator in the right hand side of
Eq.~(\ref{eqn:42}) has a zero for $\kappa b = 0$. Expanding 
the left and right hand sides of Eq.~(\ref{eqn:42}) with respect
to $k/\kappa \ll 1$ and $\kappa b \ll 1$, one gets
\begin{equation}
\label{eqn:43}
\frac{k}{\kappa} \approx \frac{kd}{2} \cdot \frac{2}{\kappa d} 
= \frac{k}{\kappa} \ll 1  
\end{equation}
Thus, Eq.~(\ref{eqn:42}) can be fulfilled well by the conditions
$k/\kappa \ll 1$ and $kd \ll 1$. Hence, the antisymmetric wave 
function is a solution of the considered task. This wave function
has consequently a zero at $x = 0$, where the potential wall is located.
The energy of the quantum state $n$ is given by $E = (\hbar k_{n})^{2}/2m$
(see Eq.~(\ref{eqn:08})). Due to the insertion of the potental wall, 
the energy of the ground state is enhanced from 
$E = (\hbar \pi/L)^{2}/2m$ (before the insertion) to
$E = 4 \times (\hbar \pi/L)^{2}/2m$ (after the insertion). 
Thus, the potential wall acts on the particle like a repulsive force.

Atkinson \& Crater \cite{atkinson73} and Busch et al. \cite{busche98}
considered the effects of an additional delta-function potential
in bound states, as for instance for a potential box and a harmonic
potential, in terms of the stationary Schr\"odinger equation. 
In the case of the potential box, as discussed here, the delta-function
was added in the middle of the box (see \cite{atkinson73}). 
Symmetric solutions with a non-vanishing value at the position of
the delta-function were found. This result contradicts to our result
presented in this Section. The reason may be that
the potential of the wall (see Eq.~(\ref{eqn:29})) does not
agree with a delta-function, since the delta-function is not actually
a function but a distribution.\footnote{Dirac's delta-function \cite{alt06} is usually given by 
\begin{eqnarray}
\label{eqn:44}
\!\!\! \delta(x) & = & \begin{cases}
                     \infty & \text{at} \quad x = 0 \\
                     0 & \text{otherwise}
                     \end{cases}
\end{eqnarray}
But, the delta-function is not actually a function but a distribution \cite{alt06} defined by
\begin{equation}
\label{eqn:45}
\int_{-\infty}^{+\infty} \dd{}x f(x) \cdot \delta(x) = f(x=0)
\end{equation}
Here, $f(x)$ is a smooth function at $x = 0$. The delta-function can be approximated by
\begin{eqnarray}
\label{eqn:46}
\!\!\! D(x) & = & \begin{cases}
                  D_{0} & \text{for} \quad -b/2 < x < b/2 \\
                  0 & \text{otherwise}
                \end{cases}
\end{eqnarray}
The function $f(x)$ can be expanded into a Taylor series around $x = 0$,
i.e. $f(x) = f_{0} + f_{1}x$ with $f_{0} = f(x=0)$ and $f_{1} = (\dd{}f/\dd{}x)_{x=0}$.
Then, one gets
\begin{equation}
\label{eqn:47}
\int_{-\infty}^{+\infty} \dd{}x f(x) \cdot D(x) = \int_{-b/2}^{b/2} f(x) \cdot D(x) 
     = f_{0} \cdot D_{0}b
\end{equation}
According to the definition (Eq.~(\ref{eqn:45})), it is justified 
to consider Eq.~(\ref{eqn:46}) as an approximation 
of the delta-function for $D_{0} \rightarrow \infty$ 
and $b \rightarrow 0$ with the condition $D_{0}b = 1$.}
\section{Classical approach of two particles in a potential box}
\label{sec:7}
A particle is considered to be trapped in a one-dimensional potential box as defined by Eq.~(2). In the framework of classical mechanics, all positions in the interval $0 < x < L$ have the same probability to be occupied by the particle. Hence, the density of probability $w_{1}$ that the particle is located in the interval $(x,x+\text{d}x)$ is constant in the box. It is given by 
\begin{eqnarray}
\label{eqn:48}
 w_{1}(x) & = & \begin{cases}
                 1/L & \text{for} \quad 0 < x < L \\
                 0   & \text{otherwise}
             \end{cases}
\end{eqnarray}
Note, that $w_{1}$ is normalized to unity, as it should be. 
Hence, the mean location of the particle is found to be
\begin{equation}
\label{eqn:49}
\bar x = \int \text{d}x \cdot x \cdot w_{1}(x) = \frac{1}{L} \cdot \int_{0}^{L} \text{d}x \cdot x = \frac{L}{2}
\end{equation}
as presumed. The mean square $\overline{(\Delta x)^{2}}$ of the fluctuations of the position of the particle around its mean value (i.e. $\bar x = L/2$) is defined by
\begin{equation}
\label{eqn:50}
\overline{(\Delta x)^{2}} = \overline{x^{2}} - (\overline{x})^{2}
\end{equation}
with
\begin{equation}
\label{eqn:51}
\overline{x^{2}} = \int \text{d}x \cdot x^{2} \cdot w_{1}(x) = \frac{L^{2}}{3}
\end{equation}
leading to
\begin{equation}
\label{eqn:52}
\overline{(\Delta x)^{2}} = \frac{L^{2}}{12}
\end{equation}
In the next step, the case of two particles trapped in the potential box is discussed. As before, all positions $x_{1}$ and $x_{2}$ in the interval $0 < x_{1},x_{2} < L$ have the same probability for both particles. Here, $x_{1}$ and $x_{2}$ denote the spatial coordinates of the \first{} and \second{} particle, respectively. Then, the density of probability that the \first{} and \second{} particle are located in the interval $(x_{1},x_{1}+\text{d}x_{1})$ and  $(x_{2},x_{2}+\text{d}x_{2})$, respectively, is given by
\begin{eqnarray}
\label{eqn:53}
w_{2}(x_{1},x_{2}) & = & \begin{cases}
                       1/L^{2} & \text{for} \quad 0 < x < L \\
                       0       & \text{otherwise}
             \end{cases}
\end{eqnarray} 
Note, that $w_{2}$ is normalized to unity. The mean distance $\bar{d}$ between both particles is defined by
\begin{equation}
\label{eqn:54}
\bar{d} = \int \text{d}x_{1} \text{d}x_{2} \cdot \|x_{1}-x_{2}\| \cdot w_{2}(x_{1},x_{2}) = \frac{L}{3}
\end{equation}
%
%
%
%
%
%
Such a value is expected, since both particles have the same distance to the walls of the potential box and to each other in the mean sense. In order to obtain the mean square $\overline{(\Delta d)^{2}}$ of the fluctuations of the distance $d$ between both particles around its mean value $\bar{d}$, the quantity
\begin{equation}
\label{eqn:55}
\overline{d^{2}} = \int \text{d}x_{1} \text{d}x_{2} \cdot \|x_{1}-x_{2}\|^{2} \cdot w_{2}(x_{1},x_{2}) = \frac{L^{2}}{6}
\end{equation}
has to be calculated. Then, one finds for
\begin{equation}
\label{eqn:56}
\overline{(\Delta d)^{2}} = \overline{d^{2}} - (\overline{d})^{2} = \frac{L^{2}}{6} - \frac{L^{2}}{9} = \frac{L^{2}}{18}
\end{equation}
%
%
%
%

%
\end{document}